\documentclass[11pt]{article}
\usepackage{moriond,epsfig}

\def\be{\begin{equation}}
\def\ee{\end{equation}}
\def\bea{\begin{eqnarray}}
\def\eea{\end{eqnarray}}

\newcommand{\pt}{\rm p_{\rm T}}
\newcommand{\ptjet}{\rm p_{\rm T}^{\rm jet}}

\newcommand{\kt}{\rm k_{\rm T}}
\newcommand{\yjet}{\rm y^{\rm jet}}

\begin{document}
\vspace*{4cm}
\title{RECENT RESULTS ON JET PHYSICS AT THE TEVATRON}

\author{ M. Mart\'\i nez}

\address{ICREA/Institut de F\'\i sica d'Altes Energies. Barcelona, 01893-E, Spain.}

\maketitle\abstracts{
In this contribution, a comprehensive review of the main aspects of high $\pt$ jet physics
in Run II at the Tevatron is presented. Recent measurements on inclusive jet and dijet production are 
discussed using different jet algorithms and covering a wide region of jet transverse momentum
and jet rapidity. Several measurements, sensitive to a proper description of soft gluon radiation
and the underlying event in hadron collisions, are also shown. 
}

\section{Inclusive Jet Production}

The measurement of the inclusive jet cross section in $p\overline{p}$ collisions at 
$\sqrt{s} = 1.96 \ \rm TeV$  constitutes a stringent test 
of perturbative QCD (pQCD) predictions over almost nine orders of magnitude.
The increased center-of-mass energy and integrated luminosity in Run II at the Tevatron 
allows to search for
signals of quark compositeness down to $\sim 10^{-19} \rm m$. 
Both CDF and D0~experiments have explored new jet algorithms away from the 
cone-based jet algorithm employed in Run I that was not infrared safe.

The CDF experiment has published
results~\cite{ktprl,runIIjet} on inclusive jet production using both the $\kt$~\cite{ktalgo,soper}  
and a  midpoint~\cite{midpoint} algorithms. While the $\kt$ algorithm is infrared safe to all orders in perturbation theory, 
the midpoint algorithm employed still suffers from some infrared sensitivity at higher orders. The measurements are  performed  for jets with
$\ptjet > 54 \ \rm GeV/c$ and rapidity  in the region $| \yjet | < 2.1$, using  
$1.0 \ \rm fb^{-1}$ of data.
\begin{figure}[h]
\mbox{\centerline{
\psfig{figure=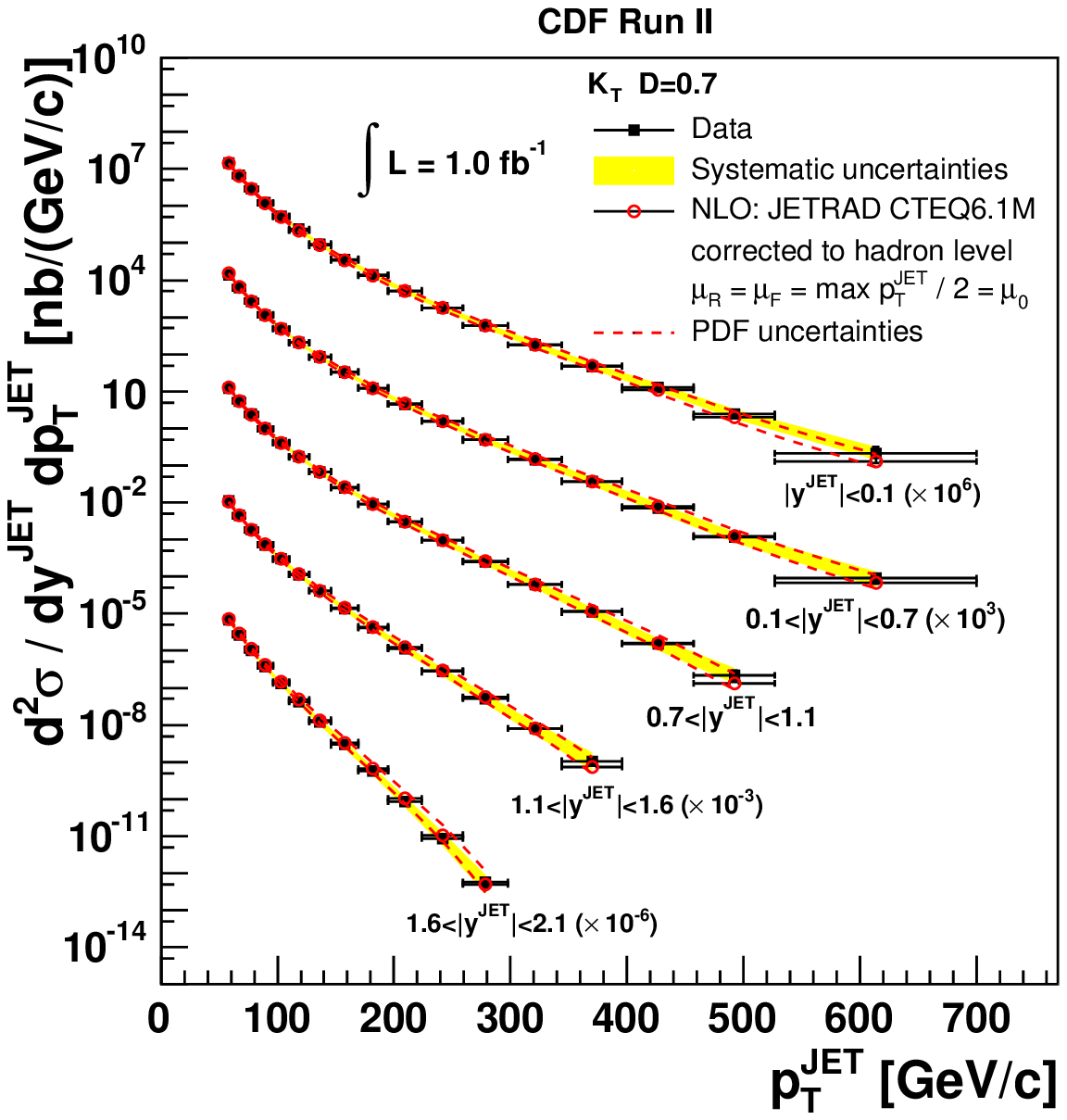,height=2.in,width=2.5in} 
\psfig{figure=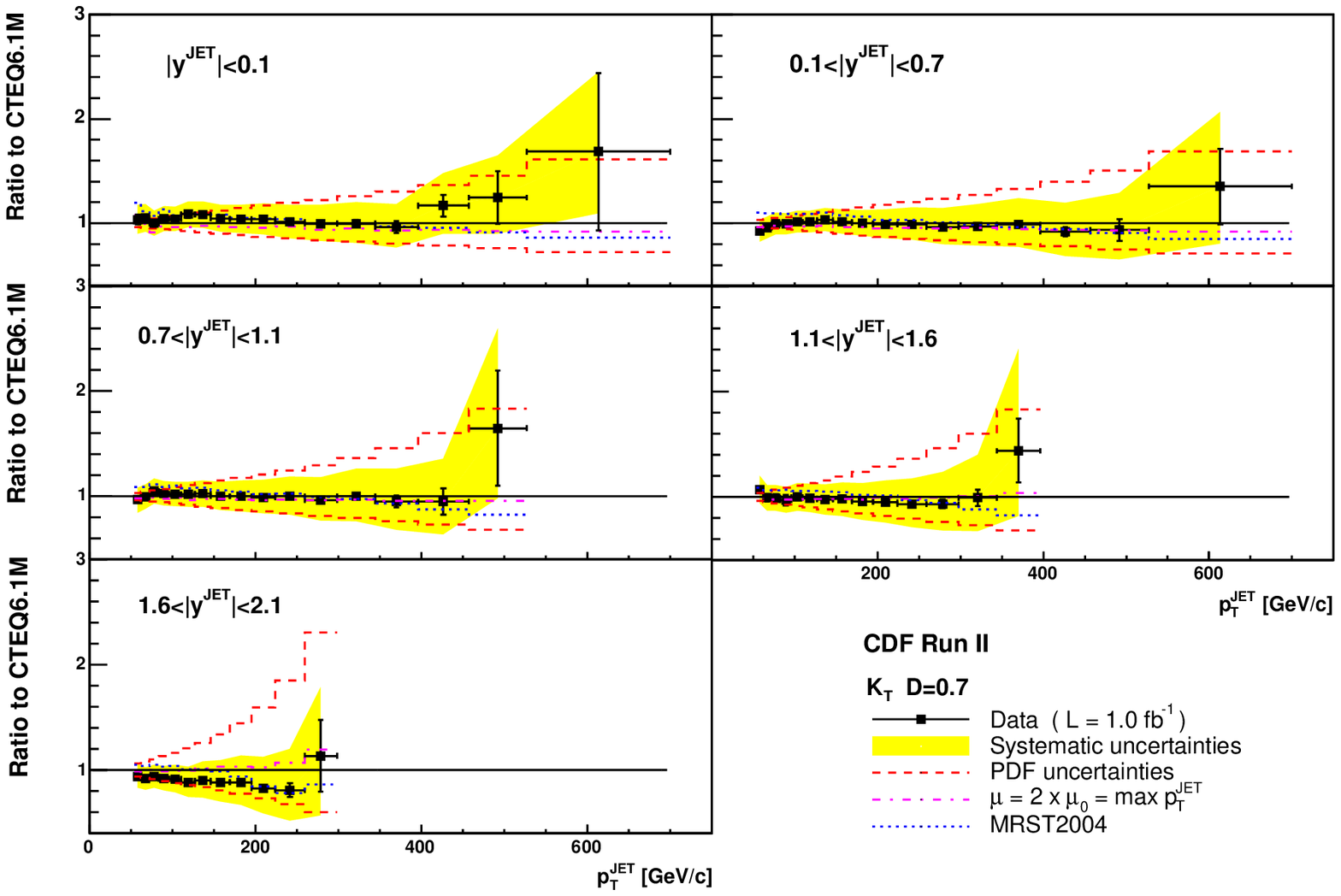,height=2.in,width=2.5in}
}}
\caption{
Measured inclusive differential jet cross sections, using the $\kt$ algorithm with $\rm D=0.7$,  as a function of $\ptjet$
in different $|\yjet|$ regions compared to NLO pQCD predictions.
The shaded bands show the total systematic uncertainty on the measurements.
The dashed lines indicate the PDF uncertainty on the theoretical predictions.}
\label{xs}
\end{figure}
Figure~\ref{xs}(left) shows the measured cross sections using the $\kt$ algorithm, with $D=0.7$,  
compared to NLO pQCD predictions~\cite{jetrad}, which include non-pQCD corrections relevant at low $\ptjet$.
The measured cross sections decrease by more than seven
to eight orders of magnitude as $\ptjet$ increases.
Figure~\ref{xs}(right) shows the ratios data/theory as a function of $\ptjet$.
Good agreement is observed in the whole range in $\ptjet$ and  $\yjet$
between the measured cross sections  and the theoretical predictions.  
In the most forward region, the uncertainty on the measured cross section  at high $\ptjet$,  compared to that on the
theoretical prediction, already indicated that the data  would contribute to
a better understanding of the gluon~PDF.
In the region $0.1 < |\yjet| < 0.7$, different values for $\rm D$ in
 the $\kt$ algorithm are considered: $\rm D=0.5$ and $\rm D=1.0$, thus decressing and increasing the effective 
size of the jet and therefore the non-pQCD contributions, respectively.
In both cases, good agreement is observed between
 the measured cross sections and the NLO pQCD predictions.

\begin{figure}
\mbox{\centerline{
\psfig{figure=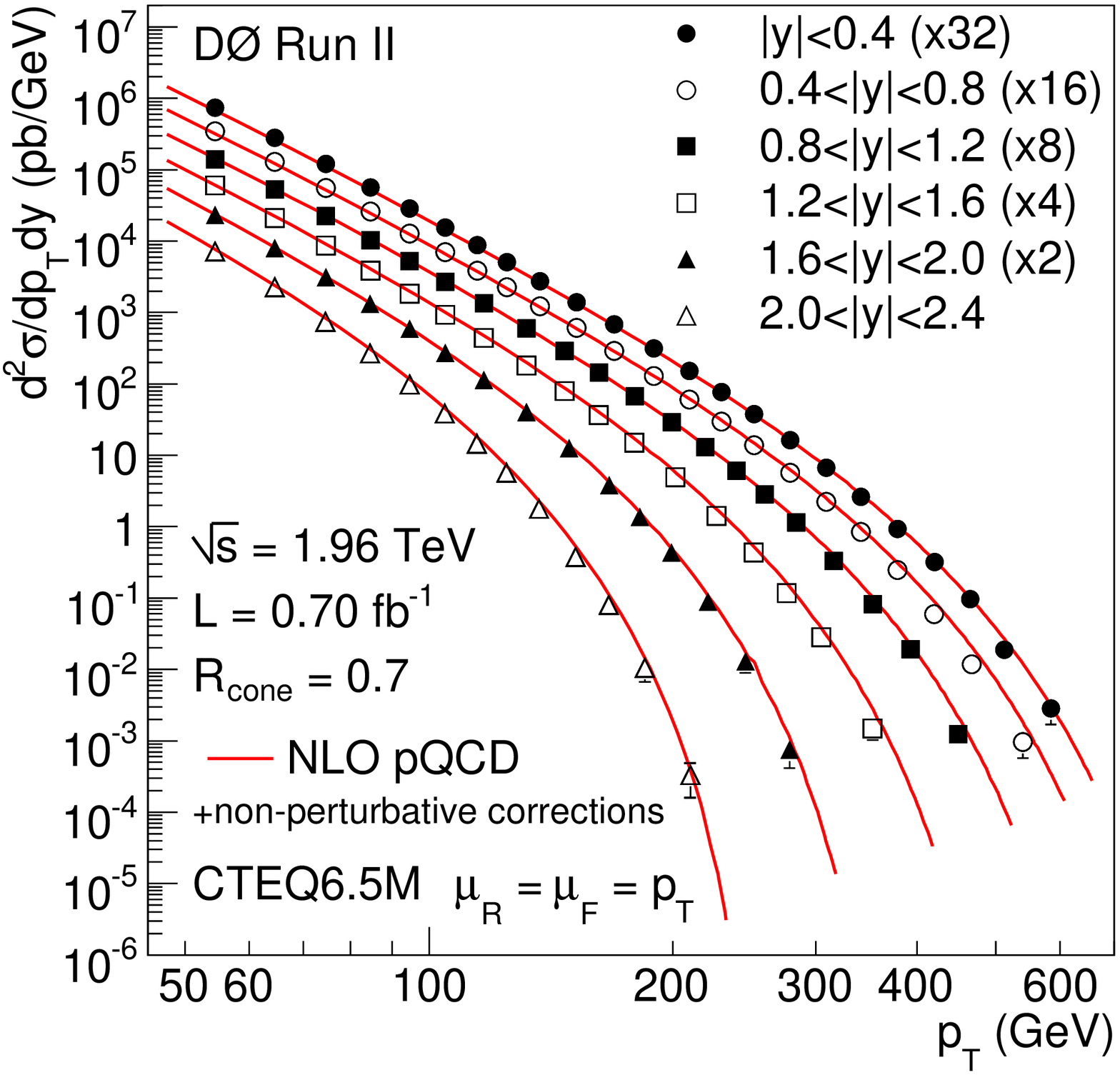,height=2.0in,width=2.5in}
\psfig{figure=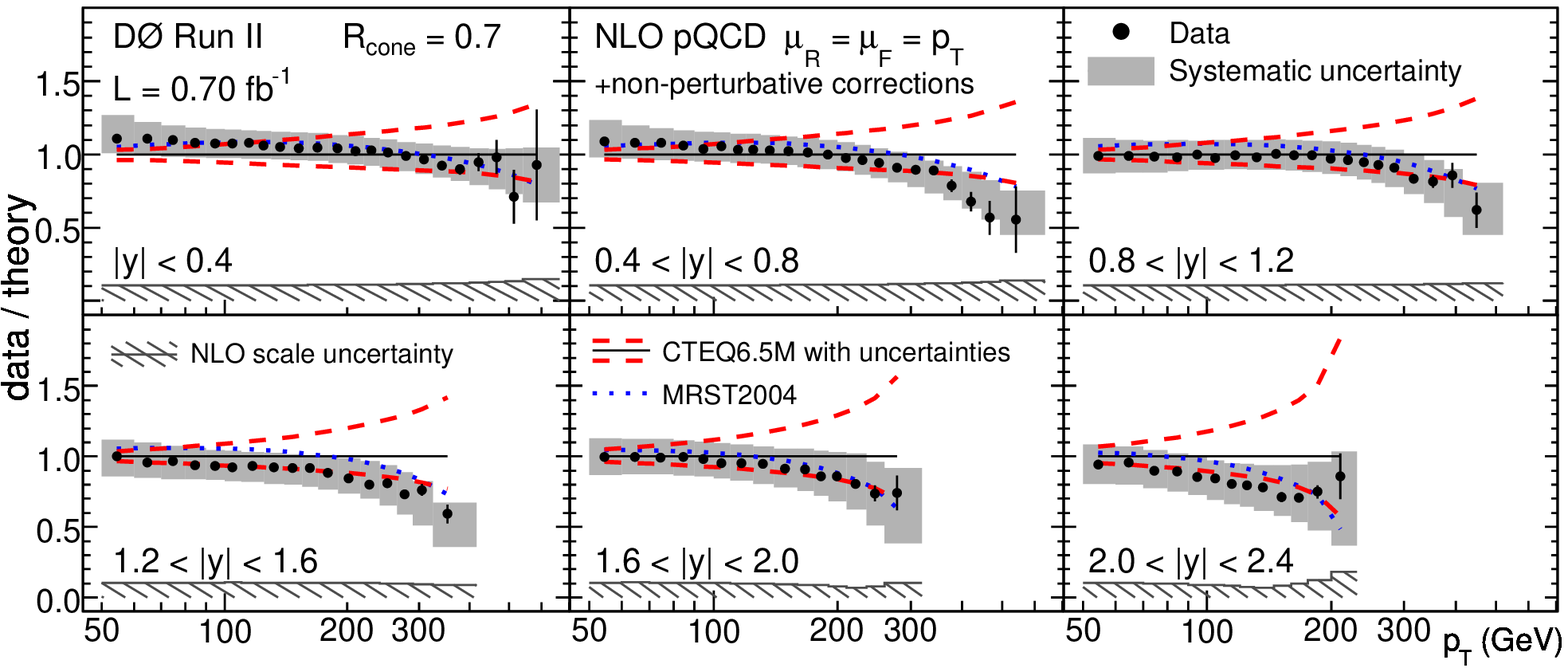,height=2.0in,width=2.5in}
}}
\caption{(left) Measured inclusive jet cross section as a function of $\ptjet$ in different $|\yjet|$ ranges 
compared to pQCD NLO predictions.(right) Ratio data/theory as a function of $\ptjet$ in different $|\yjet|$ ranges. The bands indicate
the uncertainty on the data and the dashed lines the uncertainty on the NLO prediction using  CTEQ6.5M PDFs.
The dotted lines show the ratio to MRST2004 PDFs. 
} 
\label{d0QCD1}
\end{figure}

Similarly, Figure~\ref{d0QCD1} shows the measured inclusive jet cross sections by D0~\cite{d0runII}
based on  0.7~${\rm fb}^{-1}$ of Run II data. The 
midpoint  jet algorithm has been used with a cone size R=0.7. The measurements are carried out for jets with 
$\ptjet > 50$~GeV/c in six different rapidity regions up to $|\yjet|<2.4$. The data are compared to NLO pQCD
predictions, as implemented in the NLO++~\cite{fastnlo} program, with CTEQ6.5M  and MRST2004  PDFs sets, and 
using $\ptjet$ as  the nominal renormalization/factorization scale.
Figure~\ref{d0QCD1}(right) presents the ratio data vs NLO pQCD predictions  as a function of $\ptjet$ 
in the different $|\yjet|$ regions.  The measurements are in good agreement with the theoretical predictions within
the current PDFs uncertainties. However, the Figure also suggests that the data prefer the lower edge of the CTEQ uncertainty band
while the measurements are in good agreement with the nominal MRST2004 prediction.


The CDF and D0 experiments have employed the dijet invariant mass distribution to search for resonances 
decaying into jets~\cite{dijet} as predicted by different models. In the case of D0, measurements of the dijet 
angular distributions are performed in different regions of the dijet invariant mass.  For 
both experiments, good agreement is observed between the data and theory, and the results are translated 
into improved limits in different models.  In particular, compositeness scales $\Lambda$ below 2.56 TeV 
are now excluded at 95$\%$ C.L.  

\section{Jet Shapes}

The internal structure of jets is dominated by  multi-gluon emissions from the 
primary final-state parton. It is sensitive to the relative quark- and gluon-jet fraction  and
receives contributions from soft-gluon initial-state radiation and beam remnant-remnant interactions.
The study of jet shapes at the Tevatron provides a stringent test of QCD predictions  and tests
the validity of the models for parton cascades and soft-gluon emissions in hadron-hadron collisions.
The CDF experiment has published results~\cite{shapes} on jet shapes for
central jets with transverse momentum in the region $37 < \ptjet < 380$~GeV,
where jets are searched for using the midpoint algorithm  and a cone size $R=0.7$.
The integrated jet shape, $\Psi(r)$, is defined as the average fraction of the
jet transverse momentum that lies inside a cone of radius $r$ concentric to the jet cone:
\begin{equation}
\Psi(r) = \frac{1}{\rm N_{jet}} \sum_{\rm jets} \frac{P_T(0,r) }{P_T(0,R)}, \ \ \ \ 0 \leq  r \leq R
\end{equation}
\noindent
where $\rm N_{\rm jet}$ denotes the number of jets. The measured jet shapes have been compared to the
predictions from {\sc pythia}-{\sc tune~a}  and {\sc herwig} Monte Carlo programs.
\begin{figure}
\mbox{\centerline{
\psfig{figure=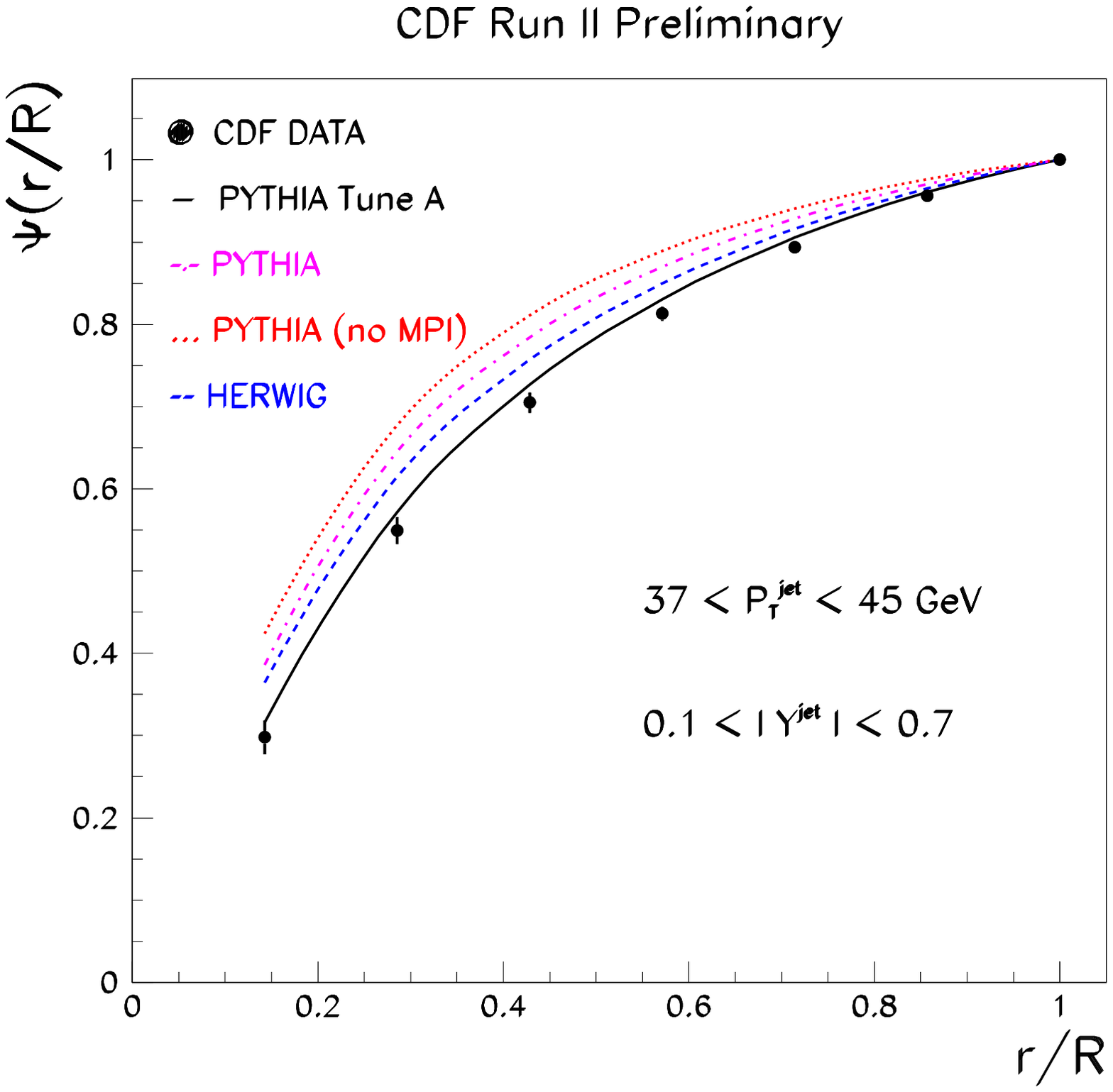,height=2.in,width=2.5in}
\psfig{figure=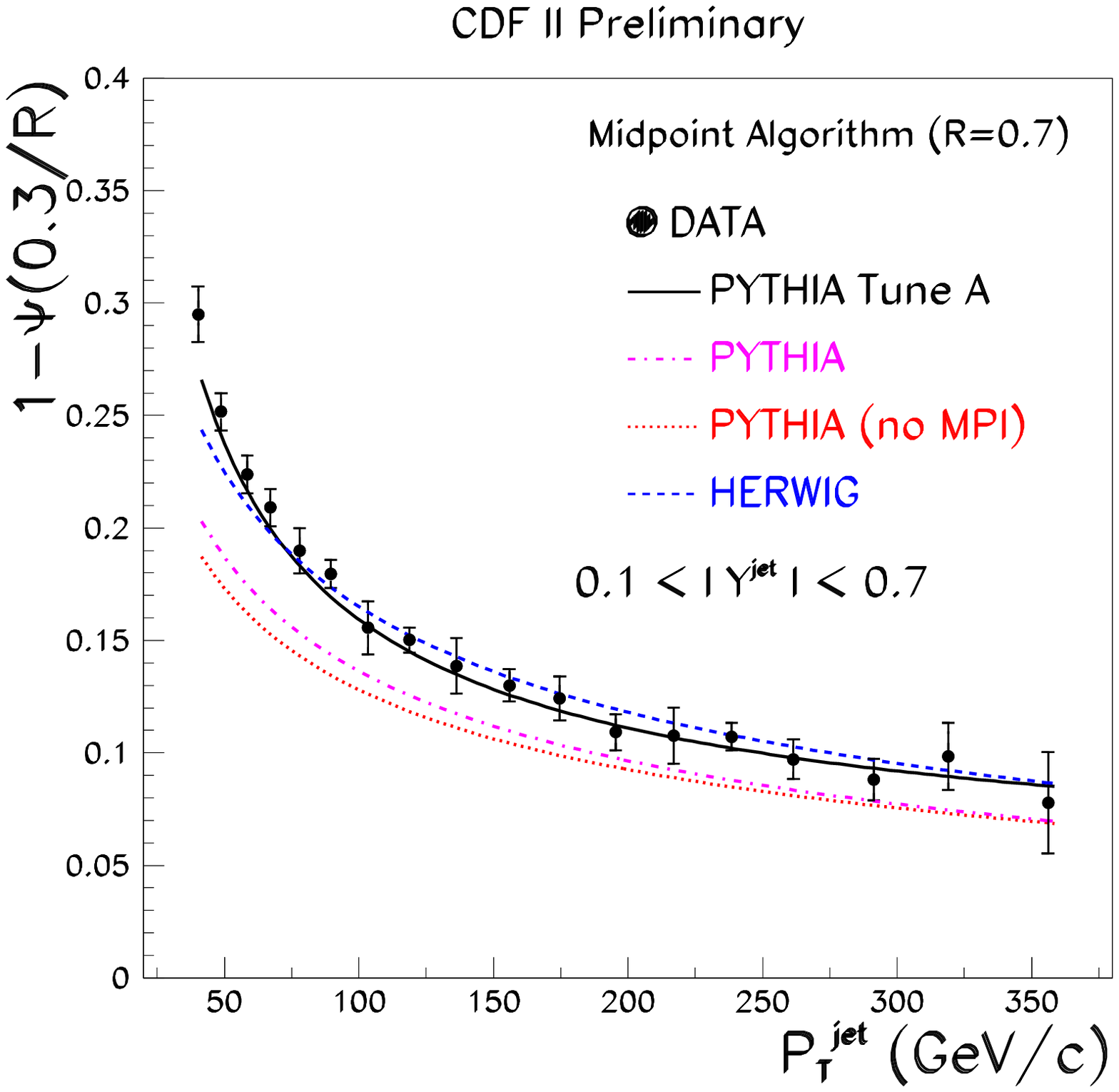,height=2.in,width=2.5in}
}}
\caption{(left) The measured integrated jet shape, $\Psi(r/R)$, in inclusive jet production for jets
with $0.1 < |\yjet| < 0.7$ and $37 \ {\rm GeV/c} < \ptjet < 45 \ {\rm GeV/c}$.  
The predictions of {\sc pythia}-{\sc tune~a}  
(solid lines), {\sc pythia} (dashed-dotted lines), {\sc pythia}-(no MPI) (dotted lines) and {\sc herwig}(dashed lines)
are shown for comparison. 
(right) The measured $1 - \Psi(0.3/R)$ as a function of $\ptjet$
for jets with $0.1 < |\yjet| < 0.7$ and $37 \ {\rm GeV/c} < \ptjet < 380 \ {\rm GeV/c}$.
} 
\label{sh1}
\end{figure}
In addition, two different {\sc pythia} samples have been used with default parameters and with and without
the contribution from multiple parton interactions (MPI) between proton and antiproton remnants, the latter
denoted as {\sc pythia}-(no MPI), to illustrate the importance of a proper modeling of soft-gluon
radiation in describing the measured jet shapes. Figure~\ref{sh1}  presents the measured integrated 
jet shapes, $\Psi(r/R)$, for jets with $37 < \ptjet < 45$ GeV, compared to
{\sc herwig}, {\sc pythia}-{\sc tune A},  {\sc pythia} and {\sc pythia}-(no MPI) predictions. 
Figure~\ref{sh1}(right)
shows, for a fixed radius $r_0 = 0.3$, the average
fraction of the jet transverse momentum outside $r=r_0$, $1-\Psi(r_0/R)$, as a function of
$\ptjet$. 
The measurements indicate
that the jets become narrower as  $\ptjet$ increases. {\sc pythia} with default parameters
produces jets systematically narrower than the data in the whole region in $\ptjet$ while 
{\sc pythia}-{\sc tune~a} predictions describe all of
the data well.

\section{Dijet Azimuthal Decorrelations}

\begin{figure}
\mbox{\centerline{
\psfig{figure=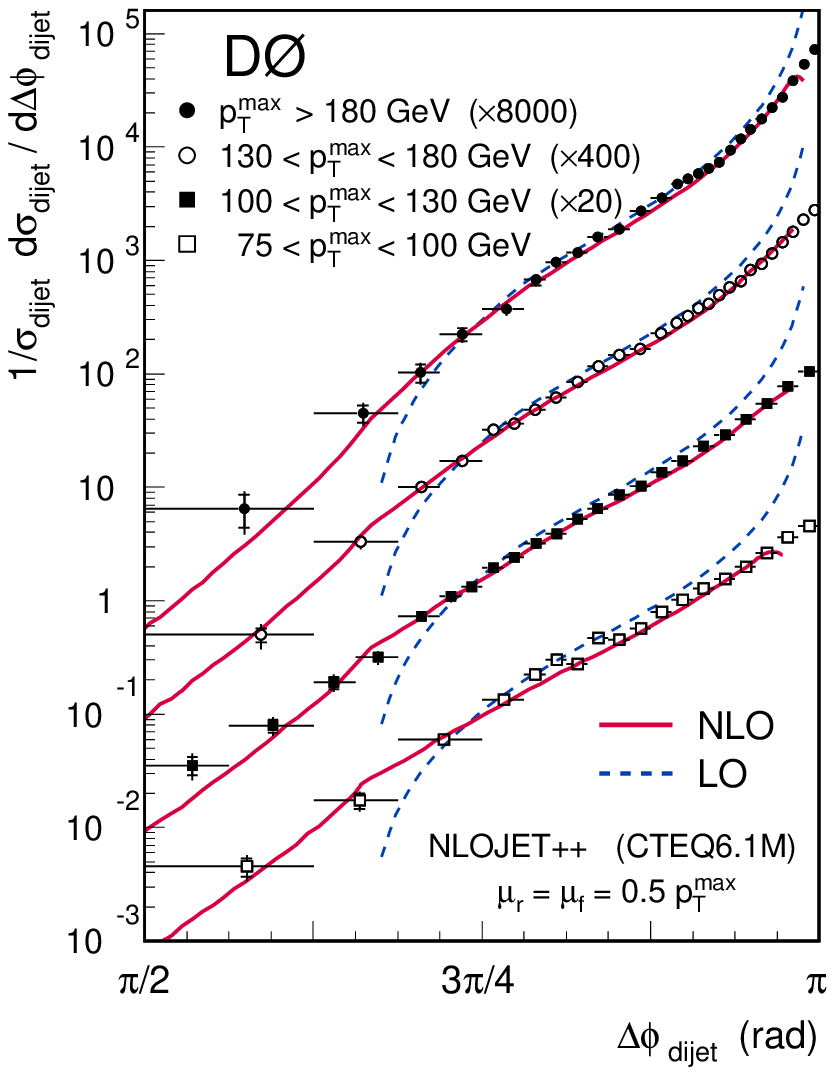,height=2.in,width=2.5in}
\psfig{figure=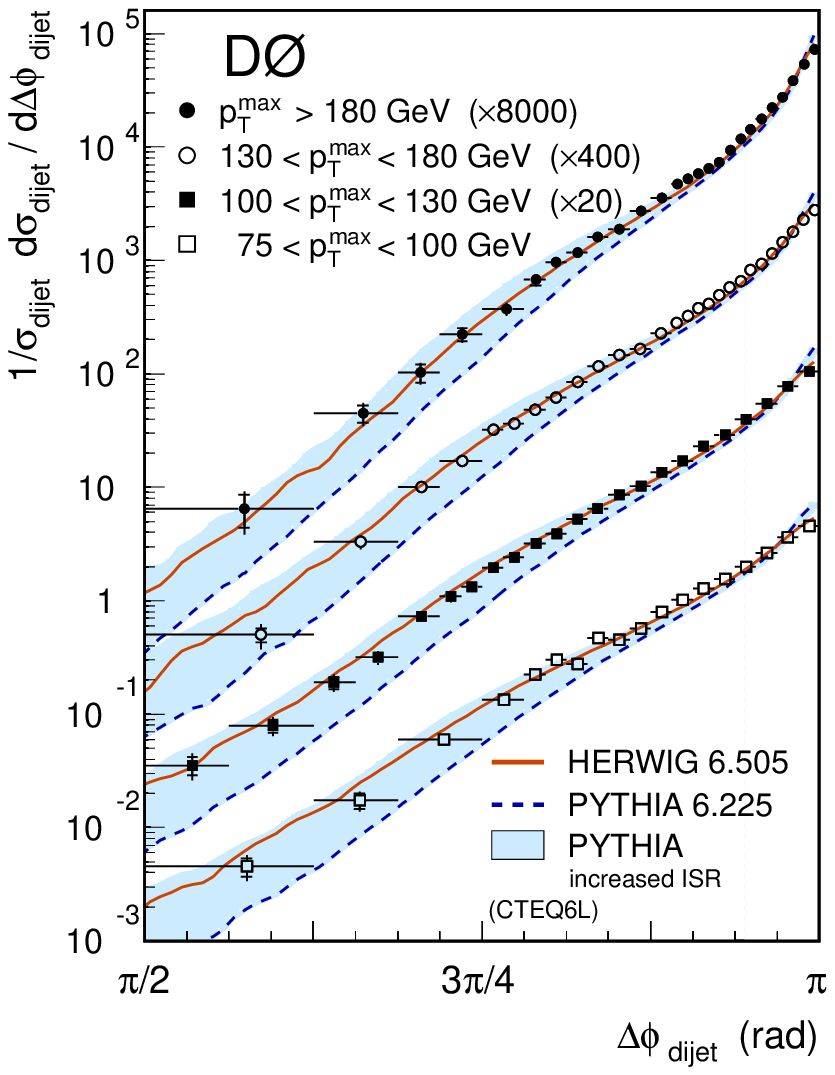,height=2.in,width=2.5in}
}}
\caption{(left) Measured azimuthal decorrelations in dijet production for central jets 
compared to pQCD predictions in
different regions of $\ptjet$ of the leading jet.
(right) Measured azimuthal decorrelations in dijet production for central jets compared 
to {\sc pythia} and {\sc herwig} predictions in different regions of leading $\ptjet$. The band covers
{\sc pythia} predictions with different amount of initial-state soft-gluon radiation.
} 
\label{dph1}
\end{figure}

The D0 experiment has employed the dijet sample to study azimuthal 
decorrelations, $\Delta \phi_{\rm dijet}$,  between the
two leading jets~\cite{dphi}. The normalized cross section,
\begin{equation}
\frac{1}{\sigma_{\rm dijet}} \frac{d \sigma}{d \Delta \phi_{\rm dijet}}, 
\end{equation}
is sensitive to the spectrum  of the gluon radiation in the event. The measurements has been performed in
different regions of the leading jet $\ptjet$ starting at $\ptjet > 75$~GeV,  where the second jet is
required to have at least $\ptjet > 40$~GeV.
Figure~\ref{dph1} shows the measured cross section compared to LO and NLO pQCD predictions~\cite{dphinlo}.
The LO (non trivial) predictions for this observable, with at most three partons in the final state, is limited to $\Delta \phi_{\rm dijet} > 2 \pi/3$,
for which the three partons define a {\it{Mercedes-star}} topology. The NLO predictions for this observable, with four partons in the final state, describes
the measured  $\Delta \phi_{\rm dijet}$ distribution better except in the very high and very low regions of  $\Delta \phi_{\rm dijet}$.  
Figure~\ref{dph1}(right) present the measured cross section compared to {\sc pythia} and {\sc herwig} predictions 
in different regions of $\ptjet$. The {\sc pythia} samples with default parameters underestimates the gluon radiation 
at large angles. Different tunes of {\sc pythia} predictions are possible, which include an enhanced contribution from initial-state 
soft gluon radiation, to properly describe the azimuthal distribution. {\sc herwig} also describes the data although 
tends to produce less radiation than {\sc pythia} close to the direction of the leading jets. 
This measurement clearly shows that angular correlations 
between jets can be employed to tune Monte Carlo predictions of 
soft gluon radiation in the final state. 

\section*{Acknowledgments}
I would like to thank organizers for their kind invitation to the conference.

\end{document}